\documentclass[12pt]{article}
\title{\bf Confining membranes and dimensional reduction}
\author{Dmitri Antonov \thanks{Tel.: +39 050 844 536; Fax: +39 050 844 538; E-mail address: {\tt antonov@df.unipi.it}}{\,}
\thanks{Permanent address:
Institute of Theoretical and Experimental Physics, 
B. Cheremushkinskaya 25, RU-117 218 Moscow, Russia. Partially supported by INTAS grant Open Call 2000, Project No. 110.}
\\
{\it INFN-Sezione di Pisa, Universit\'a degli studi di Pisa,}\\
{\it Dipartimento di Fisica, Via Buonarroti, 2 - Ed. B - I-56127 Pisa, Italy}}
\date{}
\begin{document}
\maketitle
\vspace{1mm}
\centerline{\bf {Abstract}}
\vspace{3mm}
\noindent
The dual theory describing the 4D Coulomb gas of point-like magnetically charged objects, 
which confines closed electric strings, is considered. The respective generalization of the 
theory of confining strings to confining membranes is further constructed. The same is done for the 
analogous $SU(3)$-inspired model. We then consider a combined 
model which confines both electric charges and closed strings. Such a model is nothing, but 
the mixture of the above-mentioned Coulomb gas with the condensate of the dual Higgs field, described 
by the dual Abelian Higgs model. 
It is demonstrated that in a certain limit of this dual Abelian Higgs model, 
the system under study undergoes naively the dimensional
reduction and becomes described by the (completely integrable) 2D sine-Gordon theory. In particular,
owing to this fact, the phase transition in such a model must be of the Berezinskii-Kosterlitz-Thouless
type, and the respective critical temperature is expressed in terms of the parameters of the 
dual Abelian Higgs model. However, it is finally discussed that the dimensional reduction is 
rigorously valid only in the strong coupling limit of the original 4D Coulomb gas. In such a limit, this reduction 
transforms the combined model to the 2D free bosonic theory.

\newpage

\section{Introduction}

It is known that in 4D there exists a confining medium which, 
contrary to the 4D compact QED, possesses the property of confinement 
at arbitrary values of the electric coupling~\cite{1}. Such a medium is a gas of 
point-like magnetically charged objects with a finite action. The lattice expression 
for the part of the partition function describing magnetic objects reads~\cite{1}:

\begin{equation}
\label{Zm}
{\cal Z}_m=\sum\limits_{\{q_x\}}^{}\exp\left[-2\pi^2g_m^2\sum\limits_{x,x'}^{}q_x\Delta^{-1}_{xx'}q_{x'}\right].
\end{equation}
Here, $\Delta$ is the 4D lattice Laplacian, $q_x$'s are the magnetic fluxes produced by these objects through 
3D hypercubes, and $g_m$ is the magnetic coupling constant of dimensionality 
$({\rm mass})^{-1}$. This model can be viewed as an analogue of compact QED, where the r\^ole of the vector 
potential is played by the antisymmetric tensor field (the so-called Kalb-Ramond~\cite{3} field). Clearly, such a 
field can couple only to a surface, rather than to a line as the ordinary vector potential does. Owing to this fact,
Wilson's criterion of confinement in the theory~(\ref{Zm}) should be formulated not for the usual Wilson loop, but for its analogue
defined at some closed 2D surface. The latter one can naturally be identified with a world sheet of a closed 
electric string, in the same way as the contour of the conventional Wilson loop is nothing, but a trajectory 
of an electrically charged point-like particle. Thus, confinement in the model~(\ref{Zm}) is
the confinement of closed electric strings, and its criterion is the volume law for the above-mentioned analogue of the 
Wilson loop. Such a volume law is a natural generalization to the case of strings
of the area law for the usual Wilson loops which correspond to point-like particles (see {\it e.g.}~\cite{dia}).

In the present paper, we shall make these ideas quantitative for the continuum version of the model~(\ref{Zm}). 
Our first aim, which will be realized in the next Section, will be the construction of a generalization 
of the theory of confining strings~\cite{2}, which confine electric charges, 
to the case of membranes. The resulting theory of confining membranes will be a theory which confines 
closed electric strings in the sense that the world volume of such a membrane is nothing, but a 3D hypersurface bounded by the 
world sheet of an electric string. Also, in the next Section, we shall generalize this theory to the $SU(3)$-inspired case.
In Section 3, we shall consider one concrete example of the model where closed electric strings exist, namely dual
Abelian Higgs model in the London limit. In particular, we shall explore there a possible dimensional reduction of a 
combined theory, which confines both electric strings and point-like charges, from 4D to 2D. Finally, the main results of the 
paper will be summarized in the Conclusions.

\section{Confining membranes}

The initial action of 
the model which describes the free Kalb-Ramond field and a plasma of point-like magnetic objects 
reads~\footnote{From now on, all the investigations will be performed in the 
Euclidean space-time.}

\begin{equation}
\label{1}
S=\frac{1}{12}\int d^4x\left(H_{\mu\nu\lambda}+H_{\mu\nu\lambda}^m\right)^2.
\end{equation}
Here, $H_{\mu\nu\lambda}=\partial_\mu h_{\nu\lambda}+\partial_\lambda h_{\mu\nu}+
\partial_\nu h_{\lambda\mu}$ is the field strength tensor of the Kalb-Ramond field
$h_{\mu\nu}$, $\left[h_{\mu\nu}\right]={\,}({\rm mass})$, 
satisfying the analogue of the Bianchi identity, $\varepsilon_{\mu\nu\lambda\rho}\partial_\mu
H_{\nu\lambda\rho}=0$. Next, $H_{\mu\nu\lambda}^m$ is the analogue of the monopole field strength 
tensor in compact QED, obeying the equation 

\begin{equation}
\label{bi}
\frac16\varepsilon_{\mu\nu\lambda\rho}\partial_\mu
H_{\nu\lambda\rho}^m=2\pi g_m\rho_{\rm gas}. 
\end{equation}
Here, $\rho_{\rm gas}(x)=\sum\limits_{a=1}^{N}q_a\delta(x-z_a)$ is the density of point-like  
objects, carrying magnetic charges $q_a$'s (in the units of $g_m$) and located at the points $z_a$'s.

The dualization of the theory~(\ref{1}) with the subsequent summation over the grand canonical ensemble
of magnetic objects with $q_a=\pm 1$ yields the action of the 4D sine-Gordon theory:

\begin{equation}
\label{sg}
S_{\rm SG}=\int d^4x\left[\frac12(\partial_\mu\varphi)^2-2\zeta\cos\left(2\pi g_m\varphi\right)\right].
\end{equation}
Here, $\zeta\propto\exp\left(-{\rm const}{\,}g_m^2\right)$ is the fugacity of a single magnetic object, 
$[\zeta]=({\rm mass})^4$, 
and the Debye mass of the dual Kalb-Ramond field $\varphi$, stemming from the expansion of the cosine,
reads $m=2\pi g_m\sqrt{2\zeta}$. 

One can further define in the theory~(\ref{1}) 
an analogue of the Wilson loop at the free Kalb-Ramond fields:

\begin{equation}
\label{2}
W_{\rm free}(\Sigma)=\exp\left(ig\int\limits_{\Sigma}^{}d\sigma_{\mu\nu}h_{\mu\nu}\right)=\exp\left(\frac{ig}{2}
\int\limits_{S}^{} dS_{\mu\nu\lambda}H_{\mu\nu\lambda}\right).
\end{equation}
Here, $\Sigma$ is the world sheet of a certain closed electric string, and the integration in the 
last integral on the R.H.S. of Eq.~(\ref{2}) is performed over some hypersurface $S$ enclosed by $\Sigma$.
Also, in Eq.~(\ref{2}), $g$ denotes the electric coupling constant, so that with $2\pi$ extracted 
explicitly on the R.H.S. of Eq.~(\ref{bi}), it obeys the following analogue of the Dirac quantization 
condition: $gg_m=n/3$. 
Here $n$ is a certain integer, and in what follows we shall for simplicity restrict
ourselves to the case $n=1$. Next, 
setting in Eq.~(\ref{1}) for a while $H_{\mu\nu\lambda}^m=0$, we get 

\begin{equation}
\label{3}
\left<W_{\rm free}(\Sigma)\right>=\exp\left[-g^2\int\limits_{\Sigma}^{} 
d\sigma_{\mu\nu}(x)\int\limits_{\Sigma}^{} d\sigma_{\mu\nu}(y)D_0(x-y)\right],
\end{equation}
where $D_0(x)\equiv\left(4\pi^2x^2\right)^{-1}$ is the massless propagator. The contribution of magnetic objects
to the analogue of the Wilson loop is defined by virtue of the last equality on the R.H.S. of Eq.~(\ref{2}) as follows:

\begin{equation}
\label{4}
\left<W_{\rm m}(\Sigma)\right>=\left<\exp\left(\frac{ig}{2}\int\limits_{S}^{} 
dS_{\mu\nu\lambda}H_{\mu\nu\lambda}^m\right)\right>_{\rm gas},
\end{equation}
Owing to Eq.~(\ref{bi}), it can be rewritten as 
$\prod\limits_{a=1}^{N}\exp\left[\frac{3igg_m}{\pi}q_a\eta(z_a)\right]$, where 

\begin{equation}
\label{solid}
\eta(x)=2\pi^2\partial_\mu^x\int
\limits_{S}^{} dS_\mu(y)D_0(x-y)
\end{equation}
with $dS_\mu\equiv(1/6)\varepsilon_{\mu\nu\lambda\rho}dS_{\nu\lambda\rho}$ is the solid
angle under which the hypersurface $S$ is seen by an observer located at the point $x$. As it has already been mentioned, 
this hypersurface is nothing, but the world volume of the confining membrane.
In particular for a closed hypersurface, $\eta(x)=2\pi^2$, and because of the Dirac quantization condition, 
$\left<W_{\rm m}(\Sigma)\right>$ becomes equal to unity. This fact proves that $\left<W_{\rm m}(\Sigma)\right>$
is independent of $S$, provided that the Dirac quantization condition holds.
Indeed, suppose that this is not the case and consider the ratio of two  $\left<W_{\rm m}(\Sigma)\right>$'s,
defined at different hypersurfaces, $S_1$ and $S_2$. Then, the result has the form of the R.H.S. of Eq.~(\ref{4}) with the 
integration performed over the closed hypersurface with the boundary $S_1\cup S_2$.
Therefore, such a ratio is equal to unity, which proves the desired hypersurface independence of  
$\left<W_{\rm m}(\Sigma)\right>$.
Clearly, such a consideration is analogous to the proof~\cite{4} of the fact that in compact QED with monopoles
quantized according to the Dirac quantization condition,
the monopole part of the Wilson loop does not depend on the shape of a surface bounded by the 
contour of the loop.

To construct the theory of confining membranes, it is useful to represent
the statistical weight corresponding to the kinetic term in the action~(\ref{sg}) as follows~\footnote{
We omit inessential normalization constants implying that those are included in the respective
integration measures.}:

$$\exp\left[-\frac12\int d^4x(\partial_\mu\varphi)^2\right]=
\int {\cal D}\rho\exp\left[-2\pi^2g_m^2\int d^4xd^4y\rho(x)D_0(x-y)\rho(y)
-2\pi g_m i\int d^4x\varphi\rho\right].$$
One can see that upon such a substitution, the field $\varphi$ starts playing the r\^ole of the 
Lagrange multiplier, which is used for the exponentiation of the $\delta$-function in the 
expression $1=\int {\cal D}\rho\delta(\rho-\rho_{\rm gas})$. Such a unity, being inserted into the 
summation over the grand canonical partition function, represents the Legendre transformation from 
the dual Kalb-Ramond field $\varphi$ to the dynamical density of magnetic objects, $\rho$.
Once the kinetic term of the field $\varphi$ is removed by the above substitution, this field
can further be integrated out in the saddle-point approximation. This yields the 
representation for the partition function ${\cal Z}=\int {\cal D}\varphi\exp\left(-S_{\rm SG}\right)$
in terms of dynamical densities: ${\cal Z}=\int {\cal D}\rho\exp\left(-S[\rho]\right)$.
The action $S[\rho]$ contains the Coulomb interaction of magnetic objects and their potential:

\begin{equation}
\label{srho}
S[\rho]=2\pi^2g_m^2\int d^4xd^4y\rho(x)D_0(x-y)\rho(y)+V[\rho].
\end{equation}
Here, the potential reads

\begin{equation}
\label{5}
V[\rho]=\int d^4x\left[\rho{\,}{\rm arcsinh}
\varrho-
2\zeta\sqrt{1+\varrho^2}\right] 
\end{equation}
and $\varrho\equiv\rho/(2\zeta)$.
In the dilute-gas approximation, $|\rho|\ll\zeta$, $V[\rho]\to\zeta\int d^4x\varrho^2$, and 
the integration over densities becomes Gaussian. This leads to the following expression  
for the generating functional of $\rho$'s in this limit:

\begin{equation}
\label{gf}
{\cal Z}[j]=\exp\left[-\zeta\int d^4xd^4yj(x)j(y)\partial^2D_m(x-y)\right].
\end{equation}
Here, $D_m(x)\equiv mK_1(m|x|)/(4\pi^2|x|)$ is the massive propagator with $K_1$ standing 
for the modified Bessel function. To get the expression for the contribution of magnetic objects to the analogue of the 
Wilson loop, 

\begin{equation}
\label{Zeta}
\left<W_{\rm m}(\Sigma)\right>={\cal Z}\left[\frac{i\eta}{\pi}\right],
\end{equation}
it is useful to employ the formula

$$
\partial_\mu\eta(x)=\pi^2\varepsilon_{\mu\nu\lambda\rho}\left[\int\limits_{\Sigma}^{}d\sigma_{\nu\lambda}(y)
\partial_\rho^xD_0(x-y)-\frac13\int\limits_{S}^{}dS_{\nu\lambda\rho}(y)\delta(x-y)\right].$$
Introducing the analogue of the full Wilson loop, $\left<W(\Sigma)\right>=\left<W_{\rm free}(\Sigma)\right>
\left<W_{\rm m}(\Sigma)\right>$, we obtain the following membrane action:

$$-\ln\left<W(\Sigma)\right>=$$

\begin{equation}
\label{6}
=4\pi^2\zeta\int\limits_{S}^{}dS_\mu(x)\int\limits_{S}^{}dS_\mu(y)D_m(x-y)+
\frac{g^2}{4}\int\limits_{\Sigma}^{}d\sigma_{\mu\nu}(x)\int\limits_{\Sigma}^{}d\sigma_{\mu\nu}(y)
\left[9D_m(x-y)-5D_0(x-y)\right].
\end{equation}
As it was proved after Eq.~(\ref{solid}), $\left<W_{\rm m}(\Sigma)\right>$ is independent of $S$. 
Analogously to the theory of confining strings, the $S$-dependence of the R.H.S. of Eq.~(\ref{6}),
which appears in the dilute gas approximation, 
becomes eliminated upon the summation over branches of the arcsinh-function standing 
in the full potential~(\ref{5}). This is the essence of correspondence between fields and membranes in the 
model under study. 

Note that all the above considerations can be 
performed in terms of the field ${\cal H}_{\mu\nu\lambda}$, $\varepsilon_{\mu\nu\lambda\rho}\partial_\mu
{\cal H}_{\nu\lambda\rho}\propto\rho$. Being rewritten through this field, the 
membrane representation of $\left<W(\Sigma)\right>$, given by Eq.~(\ref{Zeta}) with the generating functional
defined {\it w.r.t.} the full action~(\ref{srho})-(\ref{5}), 
is nothing, but a direct generalization of the 
theory of confining strings. Indeed, the latter one describes the string representation of the Wilson 
loop in 3D compact QED in the language of the Kalb-Ramond field $h_{\mu\nu}$, $\varepsilon_{\mu\nu\lambda}
\partial_\mu h_{\nu\lambda}\propto\rho$, where $\rho$ is the dynamical monopole density.

Next, we see that the first term on the R.H.S. of Eq.~(\ref{6}) is responsible for confinement of closed strings. 
One can parametrize the world volume of the membrane, $S$, by the vector $x_\mu(\xi)$, 
$\xi=\left(\xi^1,\xi^2,\xi^3\right)$, and expand the above-mentioned term in powers of the derivatives 
$\partial_a\equiv\partial/\partial\xi^a$ supplied with the appropriate powers of $m$ in the denominator. 
Such an expansion is similar to the analogous expansion of non-local string effective actions~\cite{exp, 2},
and its first three leading terms have the following form:

\begin{equation}
\label{7}
\zeta\int\limits_{S}^{}dS_\mu(x)\int\limits_{S}^{}dS_\mu(y)D_m(x-y)\simeq\sigma\int d^3\xi\sqrt{\hat g}+
\alpha\int d^3\xi\sqrt{\hat g}\left(K^{ia}{\,}_a\right)^2+
\beta\int d^3\xi\sqrt{\hat g}{\cal R}.
\end{equation}
Here, $\hat g^{ab}(\xi)=(\partial^ax_\mu(\xi))(\partial^bx_\mu(\xi))$ is the induced metric tensor on $S$,
$\hat g\equiv \det||\hat g^{ab}||$, and $K^{ia}{\,}_b$ is the second fundamental form of $S$ defined 
according to the Gauss-Weingarten formula:

$$D_aD_bx_\mu(\xi)=\left(\partial_a\partial_b-\Gamma_{ab}^c\partial_c\right)x_\mu(\xi)=K^i_{ab}n_\mu^i.$$
Here, $\Gamma_{ab}^c$ denotes the Christoffel symbol corresponding to the metric $\hat g^{ab}(\xi)$, and $n_\mu^i$'s, 
$i=1,2,3$, are the unit normal vectors to $S$: $n_\mu^i\partial_ax_\mu(\xi)=0$, $n_\mu^in_\mu^j=\delta^{ij}$.
Also in Eq.~(\ref{7}), 
${\cal R}$ stands the scalar curvature corresponding to the metric $\hat g^{ab}(\xi)$, which is related 
to the second fundamental form as ${\cal R}=\left(K^{ia}{\,}_a\right)^2-K^{ia}{\,}_bK^{ib}{\,}_a$.
As far as the coupling constants on the R.H.S. of Eq.~(\ref{7}) are concerned, those up to certain  
constant factors have the form:

$$\sigma\propto\frac{\zeta}{m^3}\int d^3zD_m\left(z^2\right)\propto 
\frac{\zeta}{m},~~ 
\alpha\propto\beta\propto\frac{\zeta}{m^5}\int d^3zz^2D_m\left(z^2\right)\propto 
\frac{\zeta}{m^3},$$
where $D_m\left(z^2\right)\equiv m^2K_1(|z|)/(4\pi^2|z|)$.
The first term on the R.H.S. of Eq.~(\ref{7}) is the standard membrane action~\cite{sma}.
It is analogous to the Nambu-Goto term in the action of confining strings 
and expresses confinement of a closed electric string in the 
sense of the volume law. As it has already been mentioned in the Introduction, 
the latter is nothing, but a generalization of the Wilson area law for point-like
particles to the case of strings. The second term in the expansion~(\ref{7}) 
describes the rigidity of a membrane~\cite{ridmem}. Note also that contrary to the case of confining strings,
the last term on the R.H.S. of Eq.~(\ref{7}) is not a full derivative, since the membrane action is 
three-dimensional.
 
The second term on the R.H.S. of Eq.~(\ref{6}) is analogous to the perimeter term for the Wilson loop
in 3D compact QED. However, the difference between the two theories is that in the model~(\ref{1}),
contrary to compact QED, the contribution of the free Kalb-Ramond fields~(\ref{3}) does not exactly cancel out  
with the respective massless contribution to $\left<W(\Sigma)\right>$ stemming from the magnetic objects.
However, we see that for small enough $\Sigma$'s, namely for those whose typical sizes do not exceed $1/m$, 
$\left<W(\Sigma)\right>\to\left<W_{\rm free}(\Sigma)\right>$, as it should be. In the opposite case
of $\Sigma$'s whose sizes are much larger than $1/m$, the first term on the R.H.S. of Eq.~(\ref{6}) 
dominates over the second one, and the regime of the system under study becomes purely confining.

The above presented considerations can straightforwardly be translated to the $SU(3)$-inspired case along with the 
lines of Ref.~\cite{su3}. At this point, the following remark is in order.
The $SU(3)$-inspired generalization of 3D compact QED~\cite{ws} can be viewed as 
a $[U(1)]^2$ gauge theory with monopoles, stemming from the 3D $SU(3)$ Georgi-Glashow model.  However, 
even in 3D, there does not exist a clear way to extend the usual $SU(2)$ Georgi-Glashow model
to a certain $SU(2)$-theory, which might lead to the continuous action~(\ref{1}). Therefore, the action~(\ref{1}) should be 
understood only as a theory yielding upon the dualization a continuous description of the lattice Coulomb gas~(\ref{Zm}). In the same sense, the 
$SU(3)$-inspired theory, we are going to discuss below, describes a continuous formulation of the $SU(3)$-version
of the lattice model~(\ref{Zm}). 

In the model under consideration, the charges of magnetic objects are distributed over the respective root lattice 
defined by the vectors

$$\vec q_1=\left(\frac12,\frac{\sqrt{3}}{2}\right),~ 
\vec q_2=(-1,0),~
\vec q_3=\left(\frac12,-\frac{\sqrt{3}}{2}\right),~
\vec q_{-i}=-\vec q_i.$$
The density $\rho_{\rm gas}$ becomes modified to $\vec\rho_{\rm gas}(x)=\sum\limits_{a=1}^{N}\vec q_{i_a}\delta(x-z_a)$, and 
accordingly the fields $h_{\mu\nu}$ (as well as its field strength tensor),
$H_{\mu\nu\lambda}^m$, and $\varphi$ become two-component vectors. The action~(\ref{sg}) takes the form

\begin{equation}
\label{sg3}
S_{\rm SG}^{SU(3)}=\int d^4x\left[\frac12\left(\partial_\mu\vec\varphi\right)^2-2\zeta\sum\limits_{i=1}^{3}\cos\left(2\pi g_m
\vec q_i\vec\varphi\right)\right],
\end{equation}
which yields the Debye mass $m_{SU(3)}=2\pi g_m\sqrt{3\zeta}$. With the replacement $\rho\to\vec\rho$,
the representation of the action in terms of 
densities of magnetic objects, Eq.~(\ref{srho}), keeps its form, where now 

$$V[\vec\rho{\,}]\equiv\sum\limits_{i=1}^{3}V\left[\rho_i\right].$$
In the latter formula, $V\left[\rho_i\right]$ is defined by Eq.~(\ref{5}) with $\rho_i$'s related to the components $\rho^1$ and $\rho^2$
of the vector $\vec\rho$ as follows (see Ref.~\cite{su3} for details): 
$\rho_1\equiv\left(\rho^1/\sqrt{3}+\rho^2\right)/\sqrt{3}$, 
$\rho_2\equiv -2\rho^1/3$, $\rho_3\equiv
\left(\rho^1/\sqrt{3}-\rho^2\right)/\sqrt{3}$. In the dilute gas approximation, $|\vec\rho{\,}|\ll\zeta$, the potential $V[\vec\rho{\,}]$
becomes quadratic, 
$V[\vec\rho{\,}]\to[1/(6\zeta)]
\int d^4x\vec\rho{\,}^2$. This yields for the generating functional ${\cal Z}$ of correlators of $\vec\rho{\,}$'s an expression of the form~(\ref{gf})
with the replacements $j\to\vec j$, $\zeta\to3\zeta/2$, and $m\to m_{SU(3)}$. 

The free part of the analogue of the Wilson loop is defined as follows [{\it cf.} Eq.~(\ref{2})]:

$$W_{\rm free}(\Sigma)\equiv\frac13{\,}{\rm tr}{\,}\exp\left(ig\int\limits_{\Sigma}^{}d\sigma_{\mu\nu}\vec h_{\mu\nu}\frac{\vec\lambda}{2}\right)=
\frac13\sum\limits_{\alpha=1}^{3}\exp\left(ig\vec\mu_\alpha\int\limits_{\Sigma}^{}d\sigma_{\mu\nu}\vec h_{\mu\nu}\right)=$$

\begin{equation}
\label{W3}
=\frac13\sum\limits_{\alpha=1}^{3}\exp\left(\frac{ig\vec\mu_\alpha}{2}
\int\limits_{S}^{} dS_{\mu\nu\lambda}\vec H_{\mu\nu\lambda}\right).
\end{equation}
In this equation, $\vec\lambda=\left(\lambda^3,\lambda^8\right)$ are the diagonal Gell-Mann matrices, and 
$\vec\mu_\alpha$'s stand for the weight vectors of the group $SU(3)$, 
$\vec\mu_1=\left(
-\frac12,\frac{1}{2\sqrt{3}}\right)$, $\vec\mu_2=
\left(\frac12,\frac{1}{2\sqrt{3}}\right)$, $\vec\mu_3=
\left(0,-\frac{1}{\sqrt{3}}\right)$. Averaging Eq.~(\ref{W3}) over the free Kalb-Ramond field $\vec h_{\mu\nu}$ we arrive at Eq.~(\ref{3})
with the replacement $g^2\to g^2/6$.

Similarly to Eq.~(\ref{4}), by virtue of the last equality in Eq.~(\ref{W3}) one can define the contribution of magnetic objects 
to the analogue of the Wilson loop:

$$\left<W_m(\Sigma)\right>\equiv
\frac13\sum\limits_{\alpha=1}^{3}\left<\exp\left(\frac{ig\vec\mu_\alpha}{2}
\int\limits_{S}^{} dS_{\mu\nu\lambda}\vec H_{\mu\nu\lambda}^m\right)\right>_{\rm gas}=
\frac13\sum\limits_{\alpha=1}^{3}\prod\limits_{a=1}^{N}\exp\left[\frac{2i}{\pi}\vec\mu_\alpha\vec q_{i_a}\eta(z_a)\right].$$
In the last equality, we have imposed the modified, $SU(3)$-one, Dirac quantization condition $gg_m=2n/3$, where $n$ is again some
integer which we have set for simplicity equal to unity. Next, since for any $\alpha$ and $i_a$, the scalar product $\vec\mu_\alpha\vec q_{i_a}$
is equal either to $0$ or to $\pm 1/2$,  
$\left<W_m(\Sigma)\right>$ for a closed hypersurface $S$ is equal to unity. 
This again proves that $\left<W_m(\Sigma)\right>$
is independent of the choice of $S$ [{\it cf.} the discussion after Eq.~(\ref{solid})].
The $SU(3)$-version of the theory of confining membranes is then given by the following expression for $\left<W_m(\Sigma)\right>$:

$$\left<W_m(\Sigma)\right>=\frac13\sum\limits_{\alpha=1}^{3}{\cal Z}\left[\frac{2i\vec\mu_\alpha}{\pi}\eta\right].$$
Here, the average in ${\cal Z}$'s is performed over the full action written in terms of $\vec\rho{\,}$'s, which was defined after Eq.~(\ref{sg3}).
Similarly to the $SU(2)$-inspired case, we obtain in the dilute gas approximation the nonlocal membrane action from the 
analogue of the full Wilson loop, $\left<W(\Sigma)\right>=\left<W_{\rm free}(\Sigma)\right>
\left<W_{\rm m}(\Sigma)\right>$. It reads

$$-\ln\left<W(\Sigma)\right>=8\pi^2\zeta\int\limits_{S}^{}dS_\mu(x)\int\limits_{S}^{}dS_\mu(y)D_{m_{SU(3)}}(x-y)+$$

\begin{equation}
\label{nonloc}
+\frac{g^2}{4}\int\limits_{\Sigma}^{}d\sigma_{\mu\nu}(x)\int\limits_{\Sigma}^{}d\sigma_{\mu\nu}(y)
\left[3D_{m_{SU(3)}}(x-y)-\frac73D_0(x-y)\right].
\end{equation}
Apparently, one can perform the same derivative expansion of the first term on the R.H.S. of Eq.~(\ref{nonloc}) as the one 
performed in the $SU(2)$-inspired case, and the result for the leading terms of such an expansion 
has the form of the R.H.S. of Eq.~(\ref{7}).
Notice also that as it should be, for small $\Sigma$'s, whose typical sizes do not exceed $1/m_{SU(3)}$, 
$\left<W(\Sigma)\right>\to\left<W_{\rm free}(\Sigma)\right>$.

\section{Dimensional reduction in the theory which confines closed strings and point-like charges}

As a concrete example of the theory with closed electric strings let us consider dual Abelian
Higgs model~\footnote{Below we consider for simplicity the $SU(2)$-inspired theory of confining membranes.}, 
where those are nothing, but the dual Abrikosov-Nielsen-Olesen strings~\cite{ano}.
This is also an example of the model which confines external point-like electric charges if there inserted any, just upon the 
formation of dual strings between those.
Let us consider this model in the London limit, {\it i.e.} the limit  
of infinitely heavy dual Higgs field, which means that the strings have infinitely thin cores.
In this limit, the action of the model reads

\begin{equation}
\label{sdahm}
S_{\rm DAHM}=\int d^4x\left[\frac14F_{\mu\nu}^2+\frac{\eta^2}{2}(\partial_\mu\theta-2\bar g_mB_\mu)^2\right].
\end{equation}
Here, $F_{\mu\nu}$ is the field strength tensor of the dual vector potential $B_\mu$, $\eta$ is the 
{\it v.e.v.} of the dual Higgs field, and $\bar g_m$ is the dimensionless magnetic coupling constant.~\footnote{The doubled 
magnetic charge of the dual Higgs field in Eq.~(\ref{sdahm}) is inspired by the analogy with the 3D (dual) Landau-Ginzburg
superconductor. In the latter case, this is nothing but the charge of the (magnetic) Cooper pair.}  
The phase of the dual Higgs field, $\theta$, can be decomposed into the sum of a single- and multivalued
parts. The latter one is unambiguously related to the vorticity tensor current
$\bar\Sigma_{\mu\nu}(x)\equiv\int\limits_{\Sigma}^{}d\sigma_{\mu\nu}(x(\sigma))\delta(x-x(\sigma))$ of the 
closed dual strings. In this formula, 
the vector $x_\mu(\sigma)$, $\sigma\equiv\left(\sigma^1,\sigma^2\right)$, parametrizes the string world
sheet $\Sigma$. The relation between the multivalued part of $\theta$ and $\bar\Sigma_{\mu\nu}$ 
can be expressed by means of the following equation obeyed by $\theta$:

\begin{equation}
\label{rel}
(\partial_\mu\partial_\nu-\partial_\nu\partial_\mu)\theta=\pi\varepsilon_{\mu\nu\lambda\rho}
\bar\Sigma_{\lambda\rho}.
\end{equation}
This correspondence enables one to derive the string representation 
(see {\it e.g.} Ref.~\cite{orl}) of the dual Abelian Higgs model. Namely, one can transform 
the action~(\ref{sdahm}) into the form 

\begin{equation}
\label{sr}
S_{\rm DAHM}=(\pi\eta)^2\int d^4xd^4y\bar\Sigma_{\mu\nu}(x)D_{m_B}(x-y)\bar\Sigma_{\mu\nu}(y),
\end{equation}
where $m_B=2\bar g_m\eta$ is the mass of the dual vector field, acquired by it due to the 
Higgs mechanism. 

In the representation~(\ref{sr}), dual strings are considered as individual ones, and to construct
the partition function one should supply this action with a certain prescription of the summation 
over string world sheets. One concrete example of such a prescription stems from the known fact~\cite{book}  
that at low temperatures, 
closed dual strings with minimal opposite winding numbers couple to each other and 
form virtual bound states, the so-called vortex loops. Since these objects are 
virtual and therefore small-sized, they can naturally be treated in the dilute-gas approximation.
Being rewritten in terms of dynamical densities $\Sigma_{\mu\nu}$'s of these loops, the partition 
function of the dual Abelian Higgs model takes the following form~\cite{ijmpa}:

\begin{equation}
\label{rd}
{\cal Z}_{\rm DAHM}=\int {\cal D}\Sigma_{\mu\nu}\exp\left\{-\left[
(\pi\eta)^2\int d^4xd^4y\Sigma_{\mu\nu}(x)D_{m_B}(x-y)\Sigma_{\mu\nu}(y)+V\left[\Lambda^2
\sqrt{\Sigma_{\mu\nu}^2}\right]\right]
\right\}.
\end{equation}
In this formula, $\Lambda=\sqrt{L/a^3}$ is an UV momentum cutoff, where $a$ and $L$ denote respectively  
typical sizes of vortex loops and characteristic distances between them in the gas.
In the dilute gas approximation under study, $a\ll L$ and $\Lambda\gg a^{-1}$.
Next, in Eq.~(\ref{rd}), the potential of vortex loops, $V$, is given by Eq.~(\ref{5})
with the replacement $\zeta\to\bar\zeta$. Here, $\bar\zeta\propto{\rm e}^{-S_0}$
denotes the fugacity of a single vortex loop, $\left[\bar\zeta\right]=({\rm mass})^4$, where $S_0$
is the action of a single loop estimated in Ref.~\cite{ijmpa}. In the dilute gas approximation
under study, $\sqrt{\Sigma_{\mu\nu}^2}\ll\bar\zeta/\Lambda^2$, the potential again becomes a quadratic
functional: $V\to\left[\Lambda^4/(4\bar\zeta)\right]\int d^4x\Sigma_{\mu\nu}^2$. 

The bilocal correlator of densities of the vortex loops can then be calculated by the evaluation of the 
respective Gaussian integral [{\it cf.} the generating functional~(\ref{gf})] and in addition by 
taking explicitly into account the closeness of these objects with the use of 
the Hodge decomposition theorem (see the last Ref. of~\cite{ijmpa} for details). In what follows,
we shall be interested in the limit when the {\it v.e.v.} of the dual Higgs field, $\eta$,
is very large. In that limit, the action of the grand canonical ensemble of vortex loops
standing in the exponent on the R.H.S. of Eq.~(\ref{rd}) becomes local and takes the form:
$c\int d^4x\Sigma_{\mu\nu}^2$, where $c\equiv\left(\pi^2/\bar g_m^2+\Lambda^4/\bar\zeta\right)/4$.~\footnote{Note that 
the term $(\pi/2\bar g_m)^2\int d^4x\Sigma_{\mu\nu}^2$ in this action, which does not depend on the particular properties
of the ensemble of vortex loops, can be obtained directly from the initial action~(\ref{sdahm}) by taking
in it the limit $\eta\to\infty$. Indeed, one has:

$$
\exp\left[-\frac{\eta^2}{2}\int d^4x\left(\partial_\mu\theta-2\bar g_mB_\mu\right)^2\right]=$$

$$
=\int {\cal D}\lambda_\mu\exp\left\{-\int d^4x\left[\frac{1}{2\eta^2}\lambda_\mu^2+i\lambda_\mu
\left(\partial_\mu\theta-2\bar g_mB_\mu\right)\right]\right\}
\stackrel{\eta\to\infty}{\longrightarrow}\delta\left(\partial_\mu\theta-2\bar g_mB_\mu\right).$$
Integrating further in the partition function over the $B_\mu$-field, one gets for the action~(\ref{sdahm}) 
at $\eta\to\infty$: $(4\bar g_m)^{-2}
\int d^4x\left[(\partial_\mu\partial_\nu-\partial_\nu\partial_\mu)\theta\right]^2$.  
By virtue of Eq.~(\ref{rel}) this yields the announced result.}
The above-mentioned bilocal correlator of densities of the vortex loops corresponding to this action reads:

\begin{equation}
\label{SigSig}
\left<\Sigma_{\mu\nu}(x)\Sigma_{\lambda\rho}(0)\right>=-\frac{1}{4c}\varepsilon_{\mu\nu\alpha\beta}
\varepsilon_{\lambda\rho\gamma\beta}\partial_\alpha\partial_\gamma D_0(x).
\end{equation}
Clearly, it respects the closeness of vortex loops, $\partial_\mu\Sigma_{\mu\nu}=0$. 
The higher $2n$-point correlators
of $\Sigma_{\mu\nu}$'s can be obtained from the bilocal one by means of the Wick's theorem. 

For our further purposes, we find it convenient to pass from the correlator~(\ref{SigSig}) to the 
bilocal correlator of ``hypersurfaces'' ``enclosed'' by the density of vortex loops, $\Sigma_{\mu\nu}(x)$.
The quotation marks here clearly mean that these ``hypersurfaces'' are some fields, $S_\mu$'s, corresponding
to $\Sigma_{\mu\nu}$'s according to the law [{\it cf.} the remark after Eq.~(\ref{pareta}) below]

\begin{equation}
\label{law}
\varepsilon_{\mu\nu\lambda\rho}\Sigma_{\lambda\rho}\to
\partial_\mu S_\nu.
\end{equation}
The bilocal correlator of such fields is then very simple and reads

\begin{equation}
\label{SS}
\left<S_\mu(x)S_\nu(0)\right>=\frac{1}{4c}\delta_{\mu\nu}D_0(x).
\end{equation}

Let us now consider the model, described by the sum of the actions~(\ref{sg}) 
and~(\ref{sdahm}) at $\eta\to\infty$, 
which confines both point-like electric charges and closed dual strings. 
The interaction of vortex loops with $H_{\mu\nu\lambda}^m$, described by Eq.~(\ref{4}),
can be rewritten in terms of the field $\varphi$ by virtue of the correspondence 

\begin{equation}
\label{Hph}
H_{\mu\nu\lambda}^m
\to-i\varepsilon_{\mu\nu\lambda\rho}\partial_\rho\varphi.
\end{equation}  
This yields for the partition function 
of such a model the following expression:

$${\cal Z}=\left<\int {\cal D}\varphi\exp\left[-S_{\rm SG}
-\frac{g}{2\pi^2}\int d^4x\varphi\partial^2\eta\right]\right>.$$
Here, the average is performed over the 
grand canonical ensemble of vortex loops, and $\eta(x)$ is the ``solid angle'' of the vortex loop, {\it i.e.}
the field related to $S_\mu$ by the equation 

\begin{equation}
\label{pareta}
\partial_\mu\eta=-2\pi^2S_\mu.
\end{equation} 

Note that the proof of Eq.~(\ref{pareta})
becomes straightforward by acting on both sides of this equation with the operator $\partial_\mu$ and further by 
performing for a while the inverse Legendre transformation from the 
dynamical ``hypersurfaces'' $S_\mu$'s to the real ones, corresponding to the gas of vortex loops.
The latter read $S_\mu^{\rm gas}(x)=\sum\limits_{a=1}^{N}n_a\int dS_\mu(x_a(\xi))\delta(x-x_a(\xi))$,
where $n_a=\pm 1$ denote the winding numbers of vortex loops. The desired proof then stems from the definition 
of the solid angle~(\ref{solid}). Clearly, the correspondences~(\ref{law}) and (\ref{Hph}) 
can be proved by means of the same inverse Legendre transformation.

Introducing further the notation $g(2\pi^2)^{-1}\partial^2\eta\equiv h$
we get from Eqs.~(\ref{SS}) and~(\ref{pareta}) the bilocal correlator of $h$'s:
$\left<h(x)h(0)\right>=A\delta(x)$, where $A\equiv g^2/(4c)$. Finally, rescaling the 
fields as $\varphi^{\rm new}=\varphi/\sqrt{A}$, $h^{\rm new}=h/\sqrt{A}$, we 
arrive at the following expression for the partition function of our model (we skip further for brevity the superscription ``new''):

\begin{equation}
\label{ps}
{\cal Z}=\left<\int {\cal D}\varphi\exp\left\{
-A\int d^4x\left[\frac12(\partial_\mu\varphi)^2-2\zeta A^{-1}\cos\left(\frac{\pi\varphi}{\sqrt{c}}\right)+
h\varphi\right]\right\}\right>_{\frac12\int d^4xh^2}.
\end{equation}
One can now (as it further be argued, naively) see that owing to the coupling of the dual Kalb-Ramond field $\varphi$ to the stochastic Gaussian 
field $h$ (analogous to the random external magnetic field in the case of Ising model), the model~(\ref{ps}) undergoes 
the dimensional reduction to two dimensions. The proof of this statement is similar to that of Ref.~\cite{ps}.
The only difference of the case under study from the one
considered in Ref.~\cite{ps} is that our potential of the field $\varphi$ has the form 

\begin{equation}
\label{poten}
{\cal V}=-2\zeta A^{-1}\cos\left(\frac{\pi\varphi}{\sqrt{c}}\right),
\end{equation} 
rather than that of the massive $\varphi^4$-theory. However, this 
difference is not crucial. Indeed, one can consider for instance the bilocal correlator of the mean field 
$\varphi_h$, a solution to the classical equation of motion in the theory~(\ref{ps}), 

\begin{equation}
\label{speq}
\partial^2\varphi_h-{\cal V}'(\varphi_h)=h,
\end{equation} 
which minimizes the action. It reads (see Ref.~\cite{ps} for details):

\begin{equation}
\label{100}
\left<\varphi_h(x_1)\varphi_h(x_2)\right>_{\frac12\int d^4xh^2}=
\int {\cal D}\varphi{\cal D}\alpha{\cal D}\bar\psi{\cal D}\psi\exp\left(-\int d^4x{\cal L}\right)
\varphi(x_1)\varphi(x_2).
\end{equation}
Here,

\begin{equation}
\label{101}
{\cal L}=\bar\psi\left[-\partial^2+{\cal V}''(\varphi)\right]\psi+\alpha\left[-\partial^2\varphi+{\cal V}'(\varphi)\right]
-\frac12\alpha^2
\end{equation}
and $\psi$, $\bar\psi$ are the fermionic fields which represent the Jacobian $\det|\delta h/\delta\varphi|$.
Note that the proof of Eqs.~(\ref{100})-(\ref{101}) is straightforward upon the introduction into the average on the L.H.S.
of Eq.~(\ref{100}) of the following unity:

$$1=\int {\cal D}h\delta\left[h-\left(\partial^2\varphi-{\cal V}'(\varphi)\right)\right]=$$

$$=\int {\cal D}\varphi{\cal D}\bar\psi{\cal D}\psi
\exp\left[-\int d^4x\bar\psi\left(-\partial^2+{\cal V}''(\varphi)\right)\psi\right]
\delta\left[h-\left(\partial^2\varphi-{\cal V}'(\varphi)\right)\right].$$
  
Next, similarly to the case of the potential of the massive $\varphi^4$-theory, it is straightforward
to prove that the Lagrangian~(\ref{101}) with the potential ${\cal V}$ of the general form 

\begin{equation}
\label{genpot}
{\cal V}=\sum\limits_{n=1}^{\infty}
c_n\varphi^{2n}
\end{equation} 
[and thus, in particular, with our potential~(\ref{poten})]
is invariant under the following supersymmetry transformation: 

\begin{equation}
\label{trans}
\delta\varphi=-\varepsilon_\mu x_\mu\psi,~ 
\delta\alpha=2\varepsilon_\mu\partial_\mu\psi,~
\delta\psi=0,~
\delta\bar\psi=\varepsilon_\mu x_\mu\alpha+2\varepsilon_\mu\partial_\mu\varphi,
\end{equation}
where $\varepsilon_\mu$ stands for an infinitesimal anticommuting vector. Also, similarly to Ref.~\cite{ps}, 
one can introduce a superfield 

$$\Phi\left(x,\theta,\bar\theta\right)=\varphi+\bar\theta\psi+\bar\psi\theta+\theta\bar\theta\alpha,$$
where the integration over the anticommuting variables $\theta$ and $\bar\theta$ is defined as 
$\int d\bar\theta d\theta\theta\bar\theta=1$. Furthermore, one can introduce the superspace Laplacian 
$D^2=\partial^2+\partial^2/(\partial\bar\theta\partial\theta)$, where derivatives act on the left.
With these definitions, it is possible to prove the following equality

\begin{equation}
\label{susy}
\int d^4x{\cal L}=\int d^4xd\bar\theta d\theta\left[-\frac12\Phi D^2\Phi+{\cal V}(\Phi)\right],
\end{equation}
valid for an arbitrary potential of the form~(\ref{genpot}). Such a proof becomes 
straightforward with the use of the known formula

$$\left(\sum\limits_{i=1}^{5}a_i\right)^n=\sum\limits_{{k_1+\ldots+k_5=n}\atop {k_i\ge 0}}^{}
\frac{n!}{k_1!\ldots k_5!}a_1^{k_1}\ldots a_5^{k_5},$$
where $a_i$'s represent the terms of $\Phi^2$, 

$$\Phi^2=\varphi^2+2\bar\theta\psi\varphi+2\bar\psi\theta\varphi+2\theta\bar\theta\varphi\alpha+
2\theta\bar\theta\bar\psi\psi.$$
After that, one should only pick up the terms surviving the integration over $\bar\theta$, $\theta$. 

Once the supersymmetric formulation of the action, Eq.~(\ref{susy}), is found, further proof of the dimensional reduction 
is identical to that of Ref.~\cite{ps}. Indeed, the action~(\ref{susy}) is manifestly invariant 
under rotations in superspace, which is alternatively 
expressed by the invariance of the Lagrangian~(\ref{101})-(\ref{genpot}) under 
the supersymmetry transformations~(\ref{trans}).  
Such rotations leave the interval in superspace, $\left(x^2+\theta\bar\theta\right)$, invariant.
The desired proof then completes by noting that for any supersymmetrically invariant function,
$f=f\left(x^2+\theta\bar\theta\right)$, the following equality holds:

$$
\int d^4xd\bar\theta d\theta f\left(x^2+\theta\bar\theta\right)=-\pi\int d^2xf\left(x^2\right).$$

The considerations presented above can straightforwardly be translated to an arbitrary $2n$-point correlator 
of $\varphi_h$'s. Thus, we conclude that all $2n$-point correlators of the field $\varphi_h$ 
in the theory~(\ref{ps}) are equal to the respective correlators of the field $\varphi$ computed 
in the same theory without the stochastic field $h$ and in two dimensions lower. This means that 
our 4D model confining point-like charges and closed dual strings should be described by the 2D sine-Gordon theory
of the scalar field. The latter one is known as an example of the completely 
integrable field theory (see {\it e.g.} Ref.~\cite{tf} and references therein).
In particular, at a certain critical temperature, the 2D Coulomb gas described by this theory undergoes
the Berezinskii-Kosterlitz-Thouless phase transition from the molecular to the plasma phase~\cite{bkt}. Such a phase transition occurs
when the mean squared separation between two objects forming a molecule becomes infinite. This corresponds to the divergency 
of the integral  

$$\left<r^2\right>\sim\int d^2x |x|^2\exp\left[-\frac{\ln(\mu|x|)}{2\pi T}\frac{\pi^2}{c}\right]\sim
\int\limits_{}^{\infty} d|x||x|^{3-\frac{\pi}{2cT}},$$ 
where $\mu$ denotes the IR cutoff. The respective critical temperature (which is clearly dimensionless, as it 
should be in 2D~\cite{bkt}), above which the integral diverges at the upper limit, reads

\begin{equation}
\label{tc}
T_{\rm cr}=\frac{\pi}{8c}=\frac{\pi}{2}\left(\frac{\pi^2}{\bar g_m}+\frac{\Lambda^4}{\bar\zeta}\right)^{-1}.
\end{equation}
Note that this temperature is clearly much smaller than unity, since the UV cutoff $\Lambda$ is large.
(Another reason for the smallness of $T_{\rm cr}$ can occur if we consider the case when $\bar g_m\ll 1$.
Such a situation apparently parallels the London limit of the dual Abelian Higgs model considered.)

However, there unfortunately exist indications that the above-discussed dimensional reduction to the full 2D sine-Gordon theory
is rather naive. The point is that the above-presented 
considerations are strictly speaking valid only provided that there exists the unique solution to the saddle-point
equation~(\ref{speq}) minimizing the respective action~\cite{Parisi}. Otherwise, a constraint, which is
necessary to be inserted into the functional integral in order to eliminate other solutions, leads to the violation of Eq.~(\ref{100}).
For our action with the potential~(\ref{poten}), the uniqueness of the saddle point can take place only under some approximation. Indeed, once 
such a solution is found, there apparently exists an infinite set of solutions following from that one upon the translations
of it by $2\sqrt{c}n$, where $n=\pm 1, \pm 2,\ldots$. Moreover, within half a period from a given solution there always exists another one.
The above-mentioned approximation then would mean the restriction 
of the range where $\varphi$ is defined, from the whole real axis to the one half of the period, $(-\sqrt{c}/2,\sqrt{c}/2)$. 
Such a restriction which obviously eliminates the above-discussed extra saddle points is a good approximation, since $c\gg1$ 
[{\it cf.} the discussion after Eq.~(\ref{tc})]. However, even this approximation does not guarantee us the uniqueness of the 
desired saddle-point solution, since even in such a range, the 4D sine-Gordon equation may have many solutions. 

The only rigorous 
statement concerning dimensional reduction in our case can be done if we approximate the sine-Gordon theory by the 
free bosonic one, {\it i.e.} set 

$${\cal V}\simeq \frac{\pi^2\zeta}{Ac}\varphi^2=\frac{4\pi^2\zeta}{g^2}\varphi^2.$$ 
This approximation is valid when $g$ is large enough, {\it i.e.} in the strong coupling limit of the theory~(\ref{1}).
Clearly, in such a case the saddle-point equation~(\ref{speq}) has the unique solution. Therefore, no additional constraints should be 
inserted into the integral over the $\varphi$-field, and the whole above discussion 
concerning dimensional reduction is valid. We thus conclude that the theory~(\ref{1}) in the strong coupling limit,
being coupled to the ensemble of vortex loops, undergoes the dimensional reduction to the 2D free bosonic theory.

\section{Conclusions}

In the present paper, we have considered the 4D model, which confines closed electric strings. This model
is a natural generalization of 3D compact QED to a certain 4D theory. Such a theory describes the free Kalb-Ramond field 
and a Coulomb gas of point-like magnetically charged 
objects with a finite action. We have defined in that theory an analogue of the Wilson loop and constructed for it the respective generalization 
of confining strings to confining membranes. From the so-obtained theory of confining membranes 
we have in particular deduced confinement of closed electric strings, which 
is expressed in terms of the volume law for the analogue of the Wilson loop. We have also presented the generalization of
this analysis to the $SU(3)$-inspired case. 

Next, we have considered one concrete example 
of the theory possessing closed electric strings, namely the dual Abelian Higgs model in the London limit. We have explored the 
combined theory, which confines both electric strings (due to the medium of point-like magnetic objects) and electric 
charges (due to the condensation of the dual Higgs field) in the limit when the {\it v.e.v.} of the dual Higgs field is 
infinitely large. In that case, the electric vortex loops of the dual Abelian Higgs model play for the dual Kalb-Ramond field
the same r\^ole as the random external Gaussian magnetic field plays for the mean field in the Ising model. Owing to this fact,
the system naively undergoes the dimensional reduction and becomes described by the 2D sine-Gordon theory. The latter one is known 
to be a completely integrable field theory, and many of its properties, such as the semiclassical spectrum, are known exactly.
Besides that, this theory describes a 2D Coulomb gas, which at a certain critical temperature passes through the 
Berezinskii-Kosterlitz-Thouless phase transition from the molecular to the plasma phase. The critical temperature 
of this transition is expressed in terms of the parameters of the dual Abelian Higgs model and turns out to be 
much smaller than unity. However, it is finally argued that the dimensional reduction is strictly speaking valid only in the 
case when the original theory of point-like magnetic objects is considered in the strong coupling limit. In that limit, the dimensional
reduction transforms the above-mentioned combined theory to a simple 2D free bosonic one.

\section*{Acknowledgments}
The author is indebted to Prof. A. Di Giacomo for many useful discussions and 
cordial hospitality, and to Prof. P. Menotti for a valuable discussion. He is also 
greatful to the whole staff of the Quantum Field Theory Division
of the University of Pisa for kind hospitality and to INFN for the financial support.

\end{document}